\documentclass[aps,prapplied,superscriptaddress,footinbib,floatfix,twocolumn]{revtex4-2}
\usepackage{graphicx}
\usepackage{multirow}
\usepackage{color}
\usepackage{amsmath}
\usepackage{amsfonts}
\usepackage{amssymb}
\usepackage{graphicx}
\graphicspath{{Figures/}}
\usepackage{epstopdf}
\usepackage{empheq}
\usepackage{float}
\usepackage{cases}
\usepackage{mathtools}
\usepackage{dcolumn}
\usepackage{bbold}
\usepackage{bm}
\usepackage[T1]{fontenc}

\usepackage[dvipsnames]{xcolor}
\usepackage[colorlinks]{hyperref}
\hypersetup{
colorlinks=True,
linkbordercolor=White,
linkcolor=BrickRed,
citecolor=Blue,	 
}

\newcommand\BM[1]{\textcolor{black}{#1}}

\begin{document}

\title{Mitigating variability in epitaxial-heterostructure-based spin-qubit devices by optimizing gate layout}

\author{Biel Martinez}
\email{biel.martinezidiaz@cea.fr}
\affiliation{Univ. Grenoble Alpes, CEA, LETI, F-38000, Grenoble, France}
\author{Silvano de Franceschi}
\affiliation{Univ. Grenoble Alpes, CEA, Grenoble INP, IRIG-PHELIQS, F-38000, Grenoble, France}
\author{Yann-Michel Niquet}
\email{yniquet@cea.fr}
\affiliation{Univ. Grenoble Alpes, CEA, IRIG-MEM-L\_Sim, F-38000, Grenoble, France}

\date{\today}

\begin{abstract}
The scalability of spin qubit devices is conditioned by qubit-to-qubit variability. Disorder in the host materials indeed affects the wave functions of the confined carriers, which leads to variations in their charge and spin properties. Charge disorder in the amorphous oxides is particularly detrimental owing to its long-range influence. Here we analyze the effects of charge traps at the semiconductor/oxide interface, which are generally believed to play a dominant role in variability. We consider multiple random distributions of these interface traps and numerically calculate their impact on the chemical potentials, detuning and tunnel coupling of two adjacent quantum dots in SiGe heterostructure. Our results highlight the beneficial screening effect of the metal gates. The surface of the heterostructure shall, therefore, be covered as much as possible by the gates in order to limit variability. We propose an alternative layout with tip-shaped gates that maximizes the coverage of the semiconductor/oxide interface and outperforms the usual planar layout in some regimes. This highlights the importance of design in the management of device-to-device variability.
\end{abstract}

\maketitle

\section{Introduction}

Spin qubits in semiconductor quantum dots (QDs) \cite{Loss1998,Burkard2023} provide a promising platform for quantum computing and simulation owing to a favorable ratio between gate operation and spin coherence times \cite{Yoneda2018,Piot2022}, as well as to their potential for large-scale integration, including the possibility of co-integration with classical electronics \cite{Schaal2019,Ruffino2022}. The most widely studied semiconductor spin qubits are encoded by electrons individually confined in, e.g., Si/SiO$_2$ \cite{Veldhorst2014,Veldhorst2015-2,Yang2020,Petit2020} or Si/SiGe \cite{Kawakami2014,Watson2018,Takeda2021,Mills2022,Philips2022,Unseld2023} heterostructures. The recent years have, however, witnessed a raising interest toward hole spin qubits in either Si/SiO$_2$ \cite{Maurand2016,Crippa2018,Camenzind2022,Geyer2022} or Ge/SiGe \cite{Watzinger2018,Hendrickx2020,Hendrickx2020-1,Froning2021,Hendrickx2021,wang2022ultrafast,Borsoi2023,Hendrickx2023} QDs. The various implementations span different trade-offs between electrical addressability and sensitivity of spins to noise and disorder \cite{Yoneda2018,Piot2022,Spence2022,klemt2023,Hendrickx2023}. Considerable progress has been made recently with, e.g., the operation of four Ge/SiGe hole spin qubits \cite{Hendrickx2021} and six Si/SiGe electron spin qubits \cite{Philips2022}, and the demonstration of spin-photon coupling \cite{Samkharadze2018,Mi2018,Borjans2020,HarveyCollard22,yu2022strong,Kang2023} as well as sweet spots with long dephasing times \cite{Bosco2021-1,Froning2021,Piot2022,Hendrickx2023}.

With the growing size of semiconductor quantum processors, disorder is expected to become a major hurdle to scalability \cite{Vandersypen2017,Vinet2018}. Charge disorder in the amorphous materials \cite{Varley2023,massai2023} can, in particular, deform the electronic wave functions and scatter their charge and spin properties \cite{Martinez2022,Kuppuswamy2022}. \BM{The fingerprints of variability are well visible in the few many-qubit experimental demonstrations \cite{Hendrickx2021,Philips2022,wang2024}.} Increasing the number of control gates per qubit is a way to compensate for this variability. This solution is, however, hardly compatible with the development of the large, densely packed, two-dimensional (2D) arrays of qubits required for the implementation of surface codes for quantum error correction \cite{Fowler2012,Terhal2015}. 

Dangling bonds ($P_b$ defects) at the Si/SiO$_2$ interface are particularly detrimental as they can trap both electrons and holes in the close vicinity of the qubits. One of the main advantages of heterostructures with buried quantum wells is that carriers are spatially separated from the semiconductor/oxide interface \cite{Sammak2018,Scappucci2020}. While this reduces the sensitivity to interface-trap disorder, a non-negligible variability can still persist due to the long-range nature of Coulomb interactions. 

It is, therefore, important to investigate the impact of charged interface traps on spin-qubit devices and provide guidelines for more resilient device designs. In this work, we address this problem by means of numerical simulations on a pair of quantum dots embedded in a 2D qubit array. We deliberately introduce variable microscopic configurations of charges all over the semiconductor/oxide interface and monitor their effect on the chemical potentials of the two dots, their detuning and the inter-dot tunnel coupling. These quantities determine the exchange interaction between the dots relevant for two-qubit quantum operations. We collect statistics over hundreds of charge distributions at the semiconductor/oxide interface. We focus our calculations on the case of holes in Ge/SiGe heterostructures, but the main conclusions shall also apply to other quantum-well heterostructures, in particular to the case of electrons in Si/SiGe QDs. 

We show that variability not only depends on the interface-trap density but also on the fractional coverage of the SiGe/oxide interface by the metal gates. Indeed, the latter can efficiently screen the electrostatic potential created by the charge traps beneath, but much less the potential from the traps in the spacers between the gates, which then play a dominant role. To alleviate the impact of these harmful traps, we introduce a ``tip gate'' layout where the thickness of the SiGe overlayer is deliberately increased to move the poorly screened interface traps farther away from the quantum well, and the surface gates are simultaneously replaced by tip-shape gates penetrating into the SiGe overlayer. This reshaping of the metal gates can yield to a significant improvement in dot-to-dot variability while preserving an efficient electrical control on the qubits. 

\section{Devices and methodology}
\label{sec:devices}

In the following, we introduce the devices we simulate, then discuss the methodology we use to extract the quantities relevant for two-qubit interactions.

\subsection{Planar and tip devices}

\begin{figure*}[t!]
    \centering
    \includegraphics[width=0.75\textwidth]{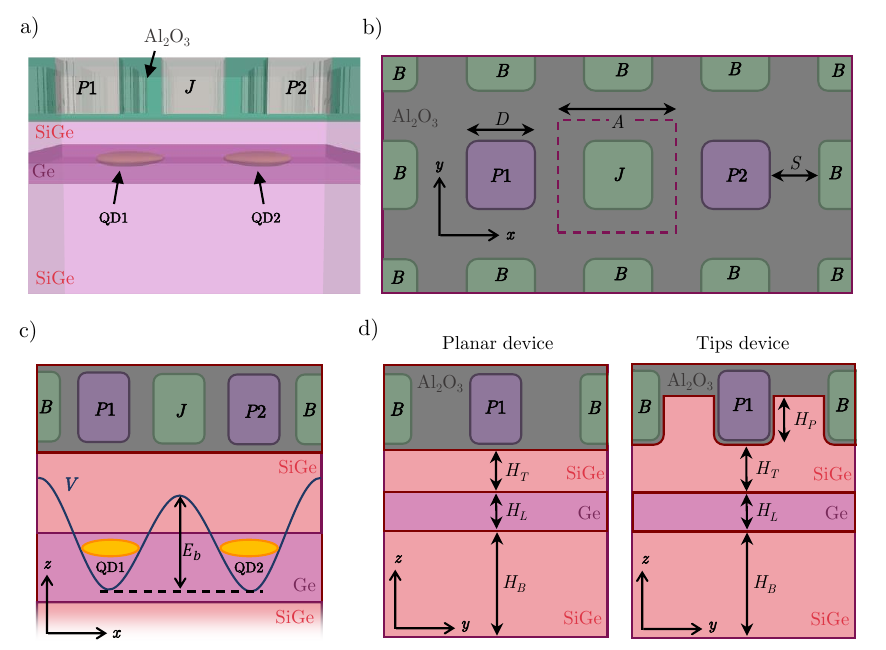}
    \caption{a) Three-dimensional representation of the simulated planar gates device. The yellow shapes are the iso-density surfaces that enclose 75\% of the hole charge below the gates. b) Top-view representation of the planar gates layout, made of rounded square plates of diameter $D$ (in green and purple) laid on the aluminium oxide (in grey). c) Cross section of the same planar gates device as in a) and b) with a sketch of the potential profile. The two dots QD$_1$ and QD$_2$ are formed with the gates $P1$ and $P2$, while the $J$ gate controls the energy barrier $E_b$, thus the tunnel coupling between the QDs (we consider holes with positive, electron-like dispersion). d) Comparison between the tip and planar gates layouts. In the latter, the gates are directly deposited on a thin oxide above the top SiGe layer. In the tip-gates layout, the top SiGe layer is grown thicker, and pits are etched and filled with a thin oxide layer and metal to form penetrating gates.}
    \label{fig:device} 
\end{figure*}

We simulate prototypical devices that emulate a 2D array of equally spaced QDs in a Ge/SiGe heterostructure, see Fig.~\ref{fig:device}. The ``planar'' gate layout consists of rounded, square metal pads deposited on a thin oxide at the surface of the heterostructure. These gates control a matrix of QDs in a Ge well beneath, and are meant to be connected with vias to upper layers of metal lines enabling electrical control. In such a versatile geometry, each gate may be operated indifferently as a plunger gate (labeled $P$ in Fig.~\ref{fig:device}) controlling the electrochemical potential of the QD underneath, or as an exchange gate (labelled $J$) controlling the inter-dot tunneling rate. For a demonstration of principle, we restrict our analysis to the two-qubit cell defined by the plunger gates $P1$ and $P2$ and exchange gate $J$ shown in Fig.~\ref{fig:device}b. We label all other gates as $B$ (barrier) gates.

In this kind of devices, the semiconductor heterostructure is usually covered by an oxide (here assumed to be Al$_2$0$_3$ as in many experiments \cite{Hendrickx2021,Hendrickx2023}). While the nature of this oxide can vary from case to case \footnote{Ge/SiGe heterostructures are often capped by a thin ($\sim 1$ nm) silicon layer that tends to partially or totally oxidize under ambient conditions. Additional oxide layers (Al$_2$O$_3$, SiO$_2$, etc...) are then deposited during the device fabrication process. In our simulations, we disregard the possible existence of the Si cap layer, and, without loss of generality, we assume that the SiGe overlayer is directly covered by an oxide.}, charge traps are expected to form at the semiconductor/oxide interface, acting as the main source of disorder \footnote{In this work, we discard other possible contributions to disorder, such as surface roughness \cite{Martinez2022,peña2023,cifuentes2023} and strain inhomogeneities due to dislocations or differential thermal contraction \cite{Abadillo2022}.} in the electrostatic potential landscape \cite{Varley2023,massai2023}. \BM{They are inherent to the amorphous nature of the oxides. Their charge state may depend on the bias and temperature history of the device \cite{Meyer2023}, and only the charged ones contribute to long-range disorder. The density $n_i$ of effectively charged traps ranges typically from $\sim 10^{11}$ to $\sim 10^{12}$ cm$^{-2}$, depending on the nature, quality, and history of the oxide \cite{massai2023}. Traps may also form within the bulk of the oxides; but they are usually fewer, and better screened by the gate (hence less harmful). They can, in a first approximation, be lumped into an effective interface trap density $n_i$ when the oxides are far enough from the dots, as is the case in Ge/GeSi heterostructures.}

In the usual case of a thin (few to several nm thick) oxide layer, the charge traps can be sorted in two classes: those lying just below the gates, and those located in the areas not covered by the gates. The former are efficiently screened by the gates, which largely mitigates their impact on the potential landscape in the Ge well. The electric field created by a charge lying at a distance $d$ from a metallic surface is, indeed, essentially dipolar at distances $r\gg d$ owing to the accumulation of an opposite image charge in the metal. Charges in the uncovered spacer areas are, however, more harmful because they are much less efficiently screened by the gates. As a result, variability is expected to decrease with gate coverage (i.e. with the $D/A$ ratio in Fig.~\ref{fig:device}b).

Following the considerations above, we explore an alternative gate geometry aimed at minimizing the effect of charge traps in the spacers. Starting from a thicker SiGe overlayer, tip-like gates are fabricated by etching holes in the SiGe and filling them with a conformal gate stack consisting of a thin oxide followed by a metal layer. In this alternative design, the harmful charges in the spacers end up lying farther away from the Ge well. While the overall surface of the SiGe/oxide interface increases by an amount corresponding to the sidewalls of the tip gates, interface traps on this additional surface have limited impact owing to the screening by the metal gates and semiconductor around. As a result, we expect improved QD uniformity, essentially limited by the closest charges trapped in the area below the gates (we implicitly assume that the interface trap density is the same around the tips and in the spacers). 

In both planar and tip devices, we form the double QD (DQD) with the plunger gates $P1$ and $P2$, and control the tunnel coupling with the exchange gate $J$. We consider a Ge/Si$_{0.2}$Ge$_{0.8}$ heterostructure where the Ge quantum well is 16 nm thick and lies 32 nm below the gates. We explore gate diameters $D$ ranging from $40$ to $70$\,nm with a fixed cell parameter $A=80$\,nm \footnote{We assume no residual strains in the Si$_{0.2}$Ge$_{0.8}$ buffer, and the strains $\varepsilon_{xx}=\varepsilon_{yy}=-0.80\%$ and $\varepsilon_{zz}=0.56\%$ are homogeneous in the Ge well.}. The gate oxide thickness (below the planar gates and around the tip gates) is set to $t_\mathrm{ox}=5$\,nm. The depth of the pits in the tip devices is $H_P=64$\,nm. We consider only positively charged traps \footnote{We note that the sign of the charge traps has little relevance when their effects can be described by first-order perturbation theory on the QD wave function \cite{Martinez2022}, as seems indeed to be the case here. Deviations with respect to the average $\sigma=\pm n_ie$ have the same statistics for both positive and negative traps.} and use a device with a homogeneous density of charges $\sigma_i=n_ie$ at the SiGe/oxide interface as a reference (accounting, therefore, for the average electrostatic potential of the traps). We ground all $B$ gates, and confine holes by imposing negative voltages to $P1$ and $P2$. We choose $V_{P1}=V_{P2}$ large enough to close the tunnel barriers with the neighboring replicas of the unit cell (residual tunneling $\tau<25$ neV). We then tune $V_J$ to achieve a target tunnel coupling $\tau=\tau^0=15$\,$\mu$eV between QD1 and QD2 (see next section for the extraction of $\tau$). Depending on the device dimensions, $V_{P1}=V_{P2}=V_P^0$ ranges from $-53$\,mV to $-72$\,mV, and $V_J=V_J^0$ ranges from $-29$\,mV to $-38$\,mV. Finally, all devices include a global back gate $128$ nm below the Ge well. The corresponding voltage is set to $0.1$\,V to mimic the small vertical electric field generated by unintentional doping in the substrate, which has anyhow a small effect. 

\subsection{Methodology}

\begin{figure}[t!]
    \centering
    \includegraphics[width=0.9\columnwidth]{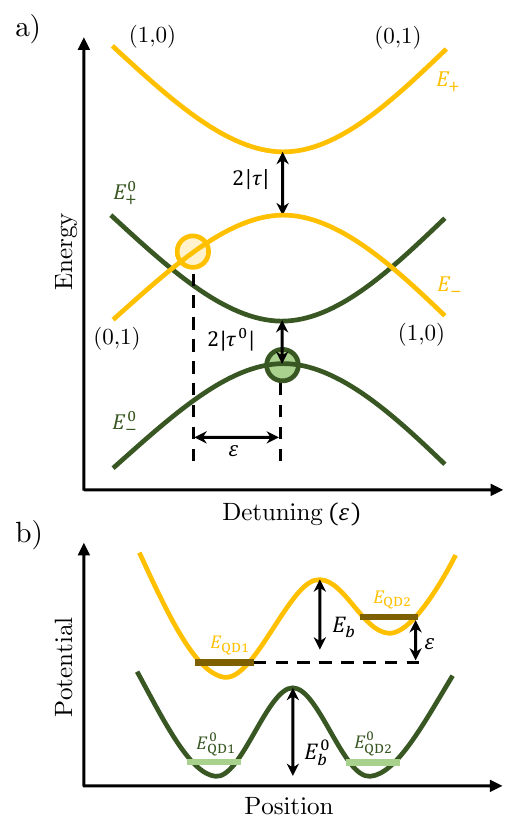}
    \caption{a) Schematic plot of the lowest-lying energy levels $E_-$ and $E_+$ as a function of detuning $\varepsilon$ in the DQD, in the absence (green lines) and presence (orange lines) of disorder. At the same bias where the pristine device is tuned ($\varepsilon=0$, green circle), the defective devices may show a finite detuning (orange circle). b) Sketch of the potential along the DQD axis for the pristine device (in green) at $\varepsilon=0$ (green circle in a)) and for a defective device (in orange) at $\varepsilon \neq 0$ (orange circle in a)). The energy levels $E_\mathrm{QD1}$ and $E_\mathrm{QD2}$ of the uncoupled quantum dots, the detuning energy $\varepsilon=E_\mathrm{QD2}-E_\mathrm{QD1}$ and the energy barrier $E_b$ are shown on this panel. When the dots are detuned, $E_b$ is measured between the top of the barrier and the average bottom of the two dots.}
    \label{fig:scheme}
\end{figure}

In this work, we specifically address the variability of single-particle quantities determining the interactions between neighboring QDs that are relevant for two qubit operations: the tunnel coupling $\tau$ and the detuning $\varepsilon=E_\mathrm{QD2}-E_\mathrm{QD1}$ between the electrochemical potentials $E_\mathrm{QD1}$ and $E_\mathrm{QD2}$ of the two QDs. Indeed, the exchange energy near the symmetric operation point is, \BM{in the simplest model}, $J=4\tau^2U/(U^2-\varepsilon^2)$, with $U$ the charging energy of the dots \cite{Burkard1999}. \BM{Although approximate, this expression emphasizes the need for a tight control of the tunnel barrier and detuning in two qubit operations.} Two-particle calculations with a configuration interaction method \cite{Abadillo2021} show that the charging energy $U$ is more robust to disorder than $\tau$ and $\varepsilon$, which shall primarily be responsible for device-to-device variations of the exchange energy (see Appendix \ref{app:U}).

Tunneling mixes the single-particle wave functions of QD1 and QD2 into a ground bonding state with energy $E_-$ and a first-excited anti-bonding state with energy $E_+$ (see Fig.~\ref{fig:scheme}):
\begin{equation}
E_\pm=\mu\pm\frac{1}{2}\sqrt{\varepsilon^2+4\tau^2}  
\label{eq:gap}
\end{equation}
with $\mu=(E_\mathrm{QD1}+E_\mathrm{QD2})/2=(E_++E_-)/2$ the average electrochemical potential of the dots. The tunnel coupling $\tau$ can therefore be extracted from the minimal gap $\Delta_\mathrm{min}=2|\tau|$ between the $E_+$ and $E_-$ branches, which occurs at $\varepsilon=0$ (i.e. for $V_{P1}=V_{P2}$ in the absence of disorder). It depends quasi-exponentially on the energy barrier $E_b$ between the dots, thus on the gate voltage $V_J$ \cite{Luyken1999}. As an illustration, $E_b^0(\tau^0=15\,\mu\mathrm{eV})=2.21$\,meV in pristine planar devices with diameter $D=50$\,nm. Increasing $E_b$ reduces $\tau$ by one decade per $2.22$\,meV (see Appendix \ref{app:tunneling}). 

The effects of disorder on the DQD are twofold: it shifts and detunes $E_\mathrm{QD1}$ and $E_\mathrm{QD2}$, and it modulates $\tau$. Therefore, we can characterize the impact of charge traps through the statistics of the electrochemical potential shift $\Delta E_\mathrm{QD}\equiv E_\mathrm{QD1}-\mu^0$ (or $E_\mathrm{QD2}-\mu^0$), of the detuning $\varepsilon$, and of the barrier height variation $\Delta E_b=E_b-E_b^0$ at the reference bias point $V_{P1}=V_{P2}=V_P^0$ and $V_J=V_J^0$ (Fig.~\ref{fig:scheme}b). We use $\Delta E_b$ rather than $\Delta\tau$ as a measure of the dispersion of tunnel couplings because it has an approximate linear dependence on potential fluctuations in the device (like $\Delta E_\mathrm{QD}$ and $\varepsilon$). Nonetheless, $\Delta E_\mathrm{QD}$, $\varepsilon$ and $\Delta E_b$ can hardly be harnessed from the data at a single bias point. Alternatively, we can characterize the impact of disorder by the bias shifts $\Delta V_{P1}=V_{P1}-V_{P1}^0$, $\Delta V_{P2}=V_{P2}-V_{P2}^0$, and $\Delta V_J=V_J-V_J^0$ needed to bring the device back to the reference conditions $E_\mathrm{QD1}=E_\mathrm{QD2}=\mu^0$ and tunnel coupling $\tau=\tau^0$. Although more dependent on the gate layout, these shifts are the actual corrections needed to operate the DQD at the nominal conditions.

To simulate the DQD devices of Figure \ref{fig:device}, we first solve Poisson's equation for the potential with a finite-volume method. We then compute the hole wave functions in this potential with a finite-difference, four-band Luttinger-Kohn (LK) model. We look for a few eigenstates of the LK hamiltonian with an iterative Jacobi-Davidson method \cite{Sleijpen1996,Sleijpen2000}. \BM{We neglect screening by the hole gases (reservoirs) far at the edges of the array and by the neighboring qubits (given the small average density of holes in the array $p<0.4\times 10^{10}$\,cm$^{-2}\ll n_i$).} We introduce the disorder at the SiGe/Al$_2$O$_3$ interface as a random distribution of positive point charges, which is fed into Poisson's equation \footnote{\BM{More precisely, the charges have an extension comparable to the mesh step. The far field (of interest in Ge/GeSi heterostructures) converges very fast with this mesh step. In Si MOS devices (see Appendix \ref{app:SiMOS}), the charges are arguably much closer to the dots and the treatment of the Coulomb singularity of the traps may require more attention. However dangling bonds ($P_b$) defects at the Si/SiO$_2$ interface capture majority carriers (electrons in $n$-type devices, holes in $p$-type devices), and thus repel the carriers in the dots. The latter do not, therefore, significantly probe the Coulomb singularity of the traps. Therefore, the calculations also converge pretty rapidly with mesh step.}}.

To bring the DQD back to $E_\mathrm{QD1}=E_\mathrm{QD2}=\mu^0$ and $\tau=\tau^0$, we need to explore the hole energy surfaces $E_-$ and $E_+$ as a function of ${\cal V}\equiv (V_{P1}, V_{P2}, V_J)$ and monitor the anti-crossing gap. Even with Newton-Raphson-type optimization algorithms, this process can be extremely slow on a finite-difference grid with $>10^6$ degrees of freedom given the large number of bias points ${\cal V}$ that need to be probed in complex potential landscapes. Therefore, we first make a rough (but fast) exploration of the hole energy surfaces around a given bias point ${\cal V}_n\equiv (V_{P1}, V_{P2}, V_J)$ by diagonalizing the Hamiltonian $H({\cal V}_n+\delta {\cal V})$ in the reduced basis set of the 64 lowest-lying eigenstates of $H({\cal V}_n)$. This yields a tentative bias point ${\cal V}_{n+1}$ where we recompute 64 new eigenstates on the finite-difference grid and iterate. As the steps $\delta {\cal V}={\cal V}_{n+1}-{\cal V}_n$ get smaller and smaller, the exploration in the reduced basis set becomes more and more accurate, expediting convergence \footnote{\BM{We consider a simulation converged when $\mu - \mu^0 < 0.01$\,meV and $\tau - \tau^0 < 0.1\,\mu$eV. Most of the devices are successfully retuned within 1-3 cycles, yet some may require up to 5-7 cycles depending on the level of disorder.}}.

\begin{table}
\begin{tabular}{l|rrr|rr|rrr}
 & $\alpha^0_{P1}$ & $\alpha^0_{P2}$ & $\alpha^0_{J}$ & $\beta^0_{P1}$ & $\beta^0_{P2}$ & $\gamma^0_{P1}$ & $\gamma^0_{P2}$ & $\gamma^0_{J}$ \\
\hline
Planar & $213.0$ & $33.9$ & $109.1$ & $-179.2$ & $179.2$ & $-58.4$ & $-58.4$ & $150.9$  \\
Tips & $221.0$ & $33.1$ & $114.5$ & $-187.9$ & $187.9$ & $-61.5$ & $-61.5$ & $158.8$ \\ 
\end{tabular}
\caption{Values (meV/V) of the $\alpha^0_i$'s, $\beta^0_i$'s and $\gamma^0_i$'s [Eqs.~\eqref{eq:leverarms}] in the planar and tip devices with $D=50$ nm and an uniform density of charges $\sigma_i=10^{11}$\,$e$/cm$^{-2}$ at the SiGe/oxide interface.}
\label{tab:leverarms_ge}
\end{table}

Finally, we estimate $\Delta E_\mathrm{QD}$, $\varepsilon$, and $\Delta E_b$ as:
\begin{subequations}
\begin{align}
\Delta E_\mathrm{QD} &= -(\alpha^0_{P1}\Delta V_{P1} + \alpha^0_{P2}\Delta V_{P2} +\alpha^0_{J} \Delta V_J) \\
\varepsilon &=-(\beta^0_{P1}\Delta V_{P1} + \beta^0_{P2}\Delta V_{P2}) \\
\Delta E_b &=-(\gamma^0_J\Delta V_J + \gamma^0_{P1}\Delta V_{P1} + \gamma^0_{P2}\Delta V_{P2})
\end{align}
\end{subequations}
where the lever arms
\begin{subequations}
\label{eq:leverarms}
\begin{align}
\alpha^0_i &= \frac{\partial E_\mathrm{QD1}}{\partial V_i} \\
\beta^0_i &= \frac{\partial (E_\mathrm{QD2}-E_\mathrm{QD1})}{\partial V_i} \\
\gamma^0_i &= \frac{\partial E_b}{\partial V_i}
\end{align}
\end{subequations}
are calculated in the reference device ($\alpha^0_i$ and $\beta^0_i$ from the slope of $E_-$ and $E_+$ at large detuning, and $\gamma^0_i$ from the total height of the barrier in the potential profile as illustrated in Fig.~\ref{fig:scheme}). These expressions translate the bias corrections needed to bring the disordered device back to nominal operating conditions into meaningful energy scales less dependent on the gate lever arms. As an illustration, the $\alpha^0_i$'s, $\beta^0_i$'s and $\gamma^0_i$'s calculated at $D=50$ nm are given in Table \ref{tab:leverarms_ge}. They are similar in the planar and tip-gate devices, and much smaller than $1$ due to the significant depth of the Ge well.

We simulate sets of 500 defective devices and quantify the variability as the standard deviation (SD) of $\Delta E_\mathrm{QD}$, $\varepsilon$ and $\Delta E_b$.

\section{Results}
\label{sec:results}

\begin{figure}[t!]
    \centering
    \includegraphics{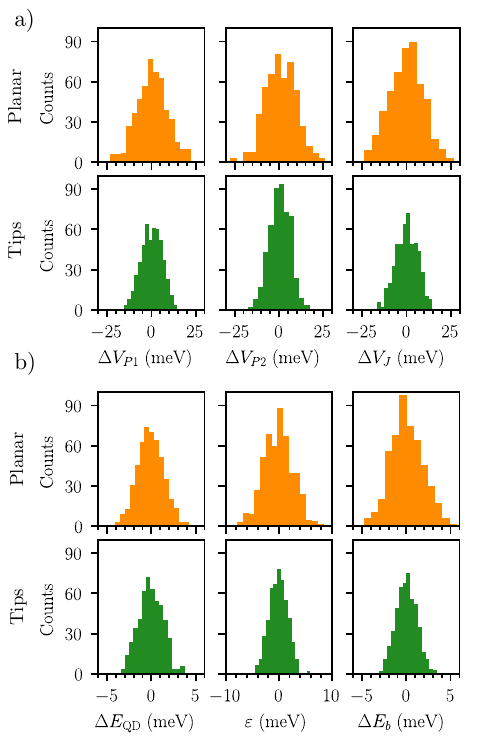}
    \caption{a) Distribution of $\Delta V_{P1}$, $\Delta V_{P2}$, and $\Delta V_J$ for 500 defective planar (orange) and tip (green) devices with gate diameter $D=50$\,nm, cell side $A=80$\,nm, and charge trap density $n_i=10^{11}$\,cm$^{-2}$.
    b) Distribution of $\Delta E_\mathrm{QD}$, $\varepsilon$, and $\Delta E_b$ in the same devices. The nominal barrier height is $E_b^0=2.21$\,meV.}
    \label{fig:counts}
\end{figure}

\begin{figure}[t!]
    \centering
    \includegraphics{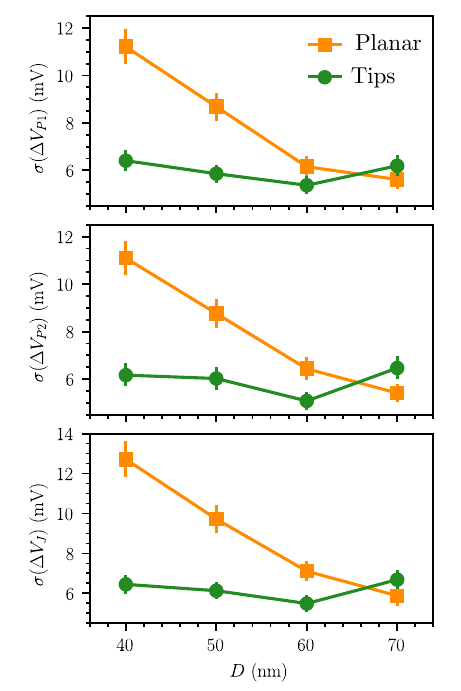}
    \caption{Dependence of the SD of $\Delta V_{P1}$, $\Delta V_{P2}$, and $\Delta V_J$ on the gate diameter $D$ for planar (orange) and tip (green) devices with cell side $A=80$\,nm and charge trap density $n_i=10^{11}$\,cm$^{-2}$.}
    \label{fig:indVvsD}
\end{figure}

\begin{figure}[t!]
    \centering
    \includegraphics{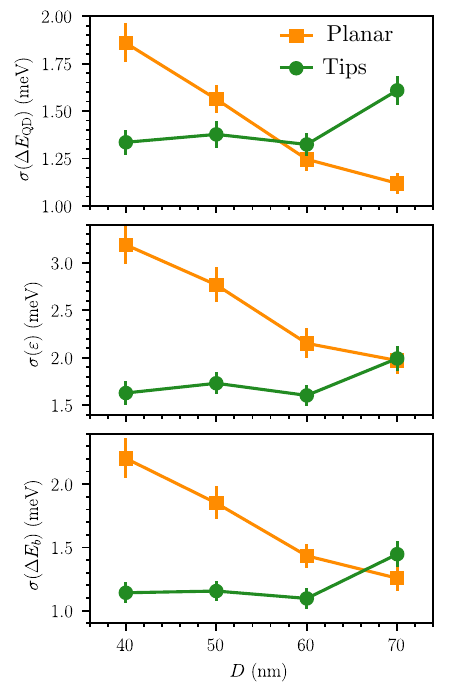}
    \caption{Dependence of the SD of $\Delta E_\mathrm{QD}$, $\varepsilon$, and $\Delta E_b$ on the gate diameter $D$ for planar (orange) and tip (green) devices with cell side $A=80$\,nm and charge trap density $n_i=10^{11}$\,cm$^{-2}$.}
    \label{fig:EvsD}
\end{figure}

We show in Fig.~\ref{fig:counts} the distributions of the bias shifts $\Delta V_{P1}$, $\Delta V_{P2}$, and $\Delta V_J$, and the energy shifts $\Delta E_\mathrm{QD}$, $\varepsilon$, and $\Delta E_b$ for a charge trap density $n_i=10^{11}$\,cm$^{-2}$. \BM{They are in all cases centered around zero, as expected from first-order perturbation theory \cite{Martinez2022}, and close to normal.} The SDs of $\Delta V_{P1}$, $\Delta V_{P2}$, and $\Delta V_J$ are plotted as a function of gate diameter $D$ in Fig.~\ref{fig:indVvsD}, and those of $\Delta E_\mathrm{QD}$, $\varepsilon$, and $\Delta E_b$ are plotted in Fig.~\ref{fig:EvsD}. Note that $\sigma(\Delta V_{P1})\approx\sigma(\Delta V_{P2})$ as expected because the two plunger gates play symmetric roles in the DQD. With a fixed cell parameter $A=80$\,nm, the gate diameter sets the gate coverage, thus the area of the weakly screened semiconductor/oxide interface. Increasing $D$ reduces this area and hence the number of unscreened charge defects. In planar gate devices, this results in a decrease of the variability, as shown quantitatively by our numerical simulations. On the other hand, we find that the variability of tip-gate devices is generally lower and rather independent of $D$. This demonstrates that the tip-gate geometry mitigates the impact of interface charge traps lying in between the gates. Only when $D\to A$, i.e. when the relative amount of these poorly screened charges becomes very small, planar-gate and tip-gate devices show similar behavior. We note that, in this limit, the potential created by the traps on the sidewalls of the tip gates gets completely suppressed by mutual screening between neighboring gates. When reducing $D$, the number of charges at the bottom and on the sidewalls of the tips decreases ($\propto D^2$ and $\propto D$, respectively), but the effective height of the tip that interacts with the DQD increases, since mutual screening subsides. These two counter-acting trends explain the weak dependence of the SDs of tip devices on the gate diameter $D$.

\begin{figure}[t!]
    \centering
    \includegraphics{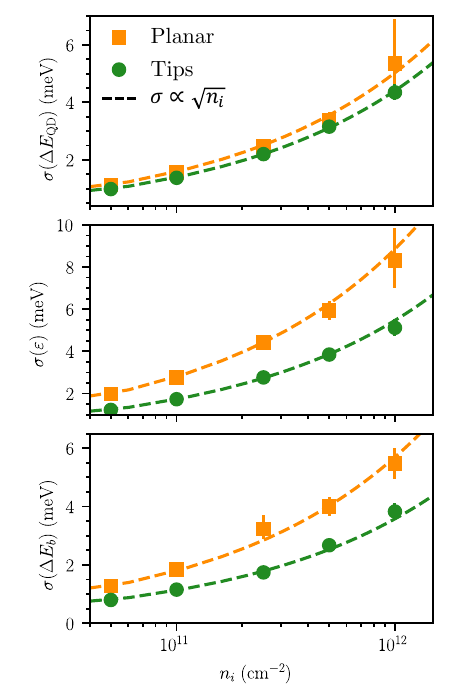}
    \caption{SD of $\Delta E_\mathrm{QD}$, $\varepsilon$, and $\Delta E_b$ as a function of the charge trap density $n_i$ for planar (orange) and tip (green) devices with gate diameter $D=50$\,nm and cell side $A=80$\,nm. Dashed lines are a guide-to-the eye for the expected $\sigma\propto \sqrt{n_i}$ dependence.}
    \label{fig:Evsdens}
\end{figure}

For planar devices, $\sigma(\Delta E_\mathrm{QD})$ reaches $1.85$\,meV at low gate coverage $D=40$ nm, which implies that $\approx 5\%$ of the dots display chemical potentials shifts $|\Delta E_\mathrm{QD}|>2\sigma(\Delta E_\mathrm{QD})=3.7$\,meV. The detuning deviation $\sigma(\varepsilon)$ is likewise greater than $3$\,meV, and the barrier height deviation $\sigma(\Delta E_b)$ is more than $2$\,meV. This large $\sigma(\Delta E_b)\gtrsim E_b$ (as well as the tail of devices with $\Delta E_b<-E_b$ in Fig.~\ref{fig:counts}b) suggests that a reasonable amount of charge disorder can suppress the tunnel barrier between the dots at bias $V_{P1}=V_{P2}=V_P^0$ and $V_J=V_J^0$. This is actually confirmed by the analysis of the potential landscape in the DQDs. Yet we find that all devices can be tuned back to nominal operating conditions. The variability of planar devices drops by almost $50\%$ when $D$ increases from 40 to 70 nm as a result of the screening of the charge traps by the metal gates. Tip-gate devices exhibit improved variability independent of $D$, yielding $\sigma(\varepsilon)\approx 1.75$\,meV and $\sigma (\Delta E_b)\approx 1.2$\,meV (these two quantities determine the mutual exchange energy). Also in this case, all devices can be tuned back to reference conditions. Practically, the variability shall remain manageable (with bias corrections) in a large array of QDs as long as $\sigma <U\simeq 3$\,meV (see Appendix \ref{app:U}).

Fig.~\ref{fig:Evsdens} shows the dependence of the variability on the charge trap density $n_i$. Improving the interface quality does reduce variability as expected. The SDs decrease by a factor of $\approx 3$ when $n_i$ is lowered from $5\times 10^{11}$ to $5\times 10^{10}$\,cm$^{-2}$. The dashed lines in Fig.~\ref{fig:Evsdens} actually emphasize the $\sigma\propto\sqrt{n_i}$ dependence expected from first-order perturbation theory \cite{Martinez2022}, and matched by the numerical results.

\section{Discussion}
\label{sec:discussion}

The results discussed above highlight that, beyond the improvement of material quality, there are two possible approaches to reduce the variability induced by interface traps in qubit devices made from epitaxial semiconductor heterostructures: either the gate coverage is maximized ($D\to A$) in order to screen the interface charges as efficiently as possible, or the charges lying in the spacers are shifted farther away from the active layer by means of tip-shaped gates. Our simulations show that both possibilities should lead to similar results. The choice is thus primarily conditioned by technical considerations. Achieving a dense two-dimensional array of planar gates requires extremely high lithographic resolution and controlled etching of the metal gate layer. Using overlapping gates can circumvent this difficulty but introduces additional oxide layers and increases the capacitive crosstalk between adjacent gates. On the other hand, the realization of tip-gate arrays requires a highly controlled etching of the semiconductor heterostructure, which can be technically challenging. The tip gates need not be very deep though, as shown in Appendix \ref{app:Hptips}. In fact, the interface traps in between the gates get almost completely screened once the tip height $H_P$ is about half of the separation between the gates. As a result, the relevant $H_P/D$ ratio is practically smaller than one, which shall ease fabrication.

\begin{figure}[t!]
    \centering
    \includegraphics{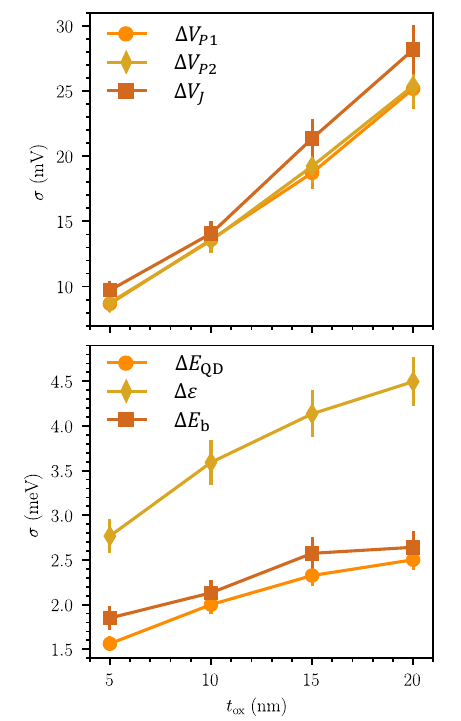}
    \caption{Dependence of the SD of the voltage corrections $\Delta V_{P1}$, $\Delta V_{P2}$, and $\Delta V_J$, and the energy shifts $\Delta E_\mathrm{QD}$, $\varepsilon$, and $\Delta E_b$ on the thickness $t_\mathrm{ox}$ of the oxide below the gates of planar devices with gate diameter $D=50$\,nm, cell side $A=80$\,nm and charge trap density $n_i=10^{11}$\,cm$^{-2}$.}
    \label{fig:tox}
\end{figure}

We emphasize that increasing the oxide thickness reduces the screening effect of the metal gates \cite{Martinez2022,Kepa23}. This is evidenced in Fig. \ref{fig:tox} for planar-gate devices. The SDs of $\Delta E_\mathrm{QD}$, $\varepsilon$ and $\Delta E_b$ raise with $t_\mathrm{ox}$ and then saturate for $t_\mathrm{ox}\gtrsim 20$ nm since the gates loose much of their screening capability. On the other hand, the SDs of $\Delta V_{P1}$, $\Delta V_{P2}$ and $\Delta V_J$ keep increasing because the gate lever arms also decrease with the oxide thickness (larger gate voltage shifts are thus needed to correct the same $\Delta E_\mathrm{QD}$, $\varepsilon$ and $\Delta E_b$). We conclude that the oxide should be as thin as possible. This can be a limiting factor in the case of qubit devices involving overlapping gate and oxide layers as opposed to single-layer ones. 

We would like to emphasize that the correction algorithm of section \ref{sec:devices} was able to tune all Ge/SiGe devices back to nominal operating conditions. We have also attempted to estimate the variability of Si/SiO$_2$ electron and hole spin qubit devices such as those of Ref.~\cite{Bedecarrats2021} (see Appendix \ref{app:SiMOS}). The charge traps, now at the Si/SiO$_2$ interface, have a much stronger impact on the qubits. With trap densities $n_i=5\times10^{10}$\,cm$^{-2}$, the variability is in fact so large that we failed to retune many devices, either because there is actually no solution, or because the algorithm of section \ref{sec:devices} is not sufficiently stable to handle strongly disordered potentials (because $\tau$ depends exponentially on the fluctuations). We expect device tuning to be similarly difficult in real experiments. This clearly underlines the need for high quality materials and interfaces, and for designs moving amorphous materials as far as possible from the active layers.

\section{Conclusions}

In this work, we have studied the impact of charge traps lying at the top SiGe/oxide interface of a Ge/SiGe spin-qubit architecture. To do so, we have considered a prototypical device consisting of a two-dimensional array of rounded square gates, and have assessed the variability of the electrochemical potentials, detuning, and interdot barrier height in a pair of adjacent quantum dots.

We have identified the ratio between gate diameter and gate pitch as a key parameter for variability. A large ratio, hence a large gate coverage of the semiconductor surface, ensures efficient screening of the interface charge defects, thus minimizing device-to-device variations. Alternatively, devices with a thicker SiGe top barrier layer and penetrating tip gates improve variability figures even for relatively small diameter/pitch ratios, since the charge defects in the metal-free regions in between the gates are shifted farther away from the underlying quantum dots. This highlights the impact of device design on variability, and provides guidelines for spin-qubit architectures more resilient to disorder, a prerequisite for scalability.

\section*{Acknowledgements}

This work was supported by the ``France 2030'' program (PEPR PRESQUILE-ANR-22-PETQ-0002) and by the European Union's Horizon 2020 research and innovation program (grant agreement 951852 QLSI).

\appendix

\section{Variability of the charging energy $U$}
\label{app:U}

\begin{figure}[ht]
    \centering
    \includegraphics{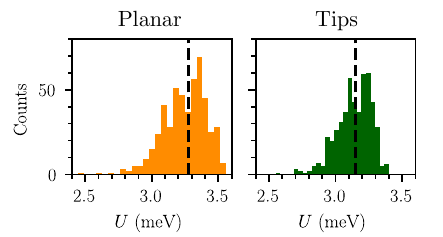}
    \caption{Distribution of charging energies $U$ for defective planar (orange) and tip (green) devices with gate diameter $D=50$ nm, cell side $A=80$\,nm, and charge trap density $n_i=10^{11}$\,cm$^{-2}$. The vertical dashed line is the charging energy for the pristine device ($U^0=3.28$ meV for the planar and $U^0=3.15$ meV for the tips device).}
    \label{fig:U} 
\end{figure}

To estimate the charging energy, we ground all gates but $P1$ in the device of Fig.~\ref{fig:device}, in order to shape a single dot. The charging energy then reads $U=E_2-2E_1$, where $E_1$ (resp. $E_2$) is the ground-state energy of the dot occupied by one (resp. two) hole(s). We compute $E_2$ with a full Configuration Interaction (CI) method \cite{Abadillo2021} in the basis set of the 48 lowest-lying single-hole states. 

The distribution of charging energies $U$ is plotted in Fig.~\ref{fig:U} for defective planar and tip devices with diameter $D=50$\,nm and $n_i=10^{11}$\,cm$^{-2}$ traps. The average charging energy is $\bar{U}=3.25$\,meV in planar and $\bar{U}=3.14$\,meV in tip devices, and the standard deviations are respectively $\sigma(U)=0.16$\,meV and $\sigma(U)=0.13$\,meV. The cost of such many-body calculations prevents, however, a systematic exploration of the variability of $U$ as a function of structural and bias parameters. The charging energy appears, nonetheless, much more robust to disorder than the tunnel coupling $\tau$ and detuning energy $\varepsilon$, which shall dominate the fluctuations of the exchange energy $J\approx 4\tau^2U/(U^2-\varepsilon^2)$.

\section{Dependence of $\tau$ and $E_b$ on $V_J$}
\label{app:tunneling}

\begin{figure}[ht]
    \centering
    \includegraphics{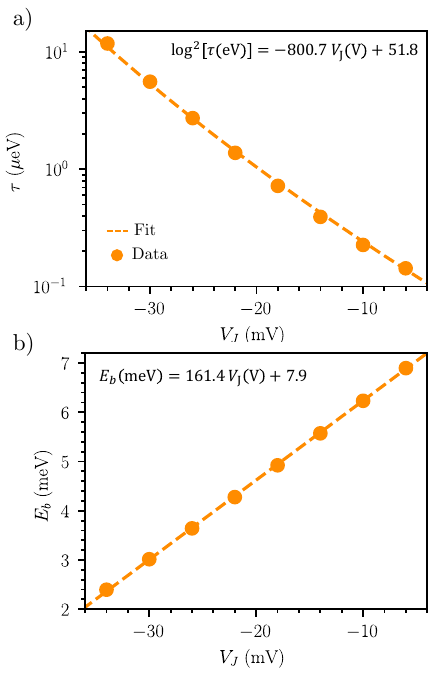}
    \caption{Dependence of (a) the tunneling coupling $\tau$ and (b) the barrier energy $E_b$ on $V_J$ for the pristine planar device with gate diameter $D=50$ nm, cell side $A=80$\,nm, and uniform charge density $\sigma_i=10^{11}$\,$e$/cm$^{-2}$ at the SiGe/Al$_2$O$_3$ interface.}
    \label{fig:tVj} 
\end{figure}

We plot in Fig.~\ref{fig:tVj} the tunnel coupling $\tau$ and the energy barrier $E_b$ as a function of the gate voltage $V_J$ for the pristine planar Ge/SiGe device with diameter $D=50$ nm (at constant $V_{P1}$ and $V_{P2}$). As expected from Wentzel–Kramers–Brillouin (WKB) approximation \cite{Luyken1999}, log$^2(\tau)$ depends almost linearly on $V_J$. The energy barrier also follows the expected linear trend.

The tunnel coupling is extracted as $\tau=\Delta_\mathrm{min}/2$, with $\Delta_\mathrm{min}$ the minimal band gap between the bonding and anti-bonding states [see Eq.~\eqref{eq:gap}]. We would like to emphasize, however, that this expression becomes inaccurate when $\tau\lesssim 500$\,neV, because the residual tunneling to the other neighboring dots $\tau^\prime\simeq 25$\,neV becomes non-negligible with respect to $\tau$. Also, the slope of $E_b(V_J)$ on Fig.~\ref{fig:tVj} is slightly different from the lever arm $\gamma_J^0$ of Table \ref{tab:leverarms}, because the former results from a fit on the whole $V_J\in[-10, -30]$\,meV range, while the latter is a single point derivative.

\section{Dependence of the variability on $H_P$ for tip devices}
\label{app:Hptips}

\begin{figure}[ht]
    \centering
    \includegraphics{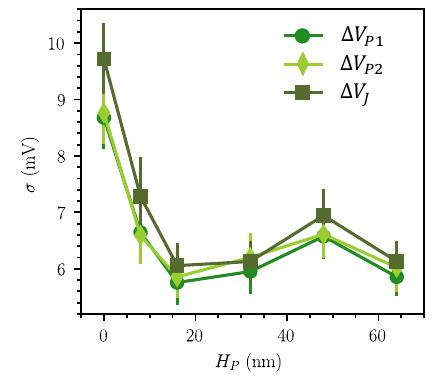}
    \caption{Dependence of the SD of $\Delta V_{P1}$, $\Delta V_{P2}$, and $\Delta V_J$ on the tip depth $H_P$ for gate diameter $D=50$\,nm, cell side $A=80$\,nm and charge trap density $n_i=10^{11}$\,cm$^{-2}$.}
    \label{fig:Hptips}
\end{figure}

In this Appendix, we assess how long the tips must be to get rid of the effect of the charge traps lying in the spacers between the gates. For that purpose, we monitor the SDs of $\Delta V_{P1}$, $\Delta V_{P2}$, and $\Delta V_J$ as a function of the depth $H_P$ of the tips (see Fig.~\ref{fig:device} for the definition of $H_P$). The data are plotted in Fig.~\ref{fig:Hptips} for a gate diameter $D=50$\,nm and for $H_P$ ranging from $0$ (the planar device) to $64$\,nm (the tip device of the main text). As expected, the variability decreases with increasing $H_P$, and levels off once $H_P\gtrsim 16$\,nm. This is, in fact, about half the distance between the gates (which are $30$\,nm apart when the cell side is $A=80$\,nm). Charges farther up are almost completely screened by the gates and do not scatter the holes significantly. Therefore, pits with small depth/diameter ratios can harvest the full benefits of the tips layout, which shall ease the development of the technology.

\section{Variability in Si MOS devices}
\label{app:SiMOS}
In this Appendix, we discuss the variability of two-qubit interactions in Si Metal-Oxide-Semiconductor (MOS) devices, for both electrons and holes.

\begin{figure}[t]
    \centering
    \includegraphics{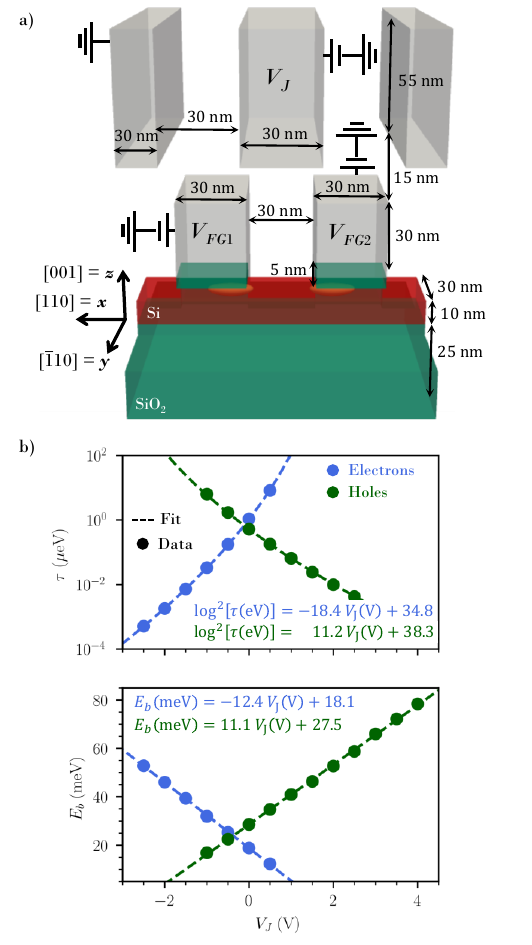}
    \caption{a) Simulated Si MOS device. A double quantum dot is formed under gates $FG1$ and $FG2$, and the $J$ gate modulates the tunnel coupling. Silicon is colored in red, metal gates in grey, and SiO$_2$ in green. The silicon oxide and silicon nitride around the channel have been removed or clarity (except under the front gates). The same device geometry is used for both electron and hole spin qubits. b) Dependence of the tunnel coupling $\tau$ and energy barrier $E_b$ on $V_J$ for pristine devices with a uniform charge density $\sigma_i=\pm 5\times 10^{10}$\,cm$^{-2}$ at the Si/SiO$_2$ interface. The fits are used to convert the $\tau$'s calculated in defective devices into $\Delta E_b$'s.}
    \label{fig:supp_device} 
\end{figure}

For that purpose, we simulate the device of Fig.~\ref{fig:supp_device}a, similar to the fully-depleted silicon-on-insulator (FDSOI) design of Ref. \cite{Bedecarrats2021}. The rectangular silicon channel is $10$\,nm thick and $30$\,nm wide and is lying on a $25$\,nm thick buried SiO$_2$ and a Si substrate used as a (grounded) back gate. The device is controlled by two levels of gates: a first one with partly overlapping front gates ($FG$) that shape the dots, and a second one with interleaved exchange gates ($J$) that control the tunnel coupling between the dots. The channel is embedded in a $5$\,nm thick SiO$_2$ gate oxide and the whole device is encapsulated in Si$_3$N$_4$. The simulated cell of Fig.~\ref{fig:supp_device}a is made periodic along the channel axis $x=[110]$. Both front gates are set to $V_{FG}=50$ mV for electrons and $V_{FG}=-50$ mV for holes and the outermost $J$ gate is biased to shut down tunneling with the neighboring cells. The central $J$ gate is then tuned to achieve a tunnel coupling $\tau^0=10$\,$\mu$eV between the two dots ($V_J=0.54$\,V for electrons and $V_J=-1.16$\,V for holes).

\begin{table*}
\begin{tabular}{l|rrr|rr|rrr}
 & $\alpha^0_{P1}$ & $\alpha^0_{P2}$ & $\alpha^0_{J}$ & $\beta^0_{P1}$ & $\beta^0_{P2}$ & $\gamma^0_{P1}$ & $\gamma^0_{P2}$ & $\gamma^0_{J}$ \\
\hline
Electrons & $-708.5$ & $-94.5$ & $-15.5$ & $-614.0$ & $614.0$ & $30.6$ & $30.6$ & $-10.5$  \\
Holes & $742.2$ & $86.4$ & $12.6$ & $-655.8$ & $655.8$ & $-34.8$ & $-34.8$ & $9.9$ \\ 
\end{tabular}
\caption{Values (meV/V) of the $\alpha^0_i$'s, $\beta^0_i$'s and $\gamma^0_i$'s (as defined by Eq. (3) of the main text) for electrons and holes in the Si MOS devices with a uniform density of charges $\sigma_i=\pm5\times10^{10}$\,$e$/cm$^{-2}$ at the Si/SiO$_2$ interface.}
\label{tab:leverarms}
\end{table*}

We introduce charge disorder at the Si/SiO$_2$ interface as $P_b$ (dangling bonds) defects that capture majority carriers \cite{Martinez2022}. We thus model them as negative (resp. positive) point charges for electrons (resp. holes) with areal density $n_i=5\times 10^{10}$ cm$^{-2}$. Even though $n_i$ is lower than in Ge/GeSi heterostructures, the traps are much closer to the dots, which strengthens their impact \cite{Martinez2022,cifuentes2023}.

\begin{figure}
    \centering
    \includegraphics{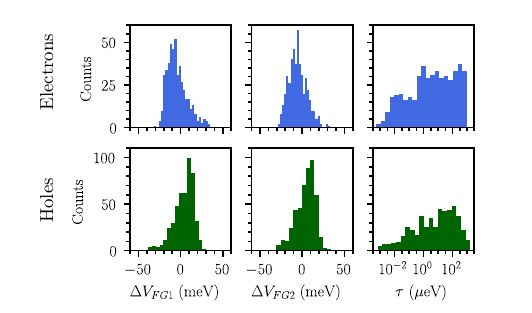}
    \caption{Distribution of gate voltage corrections $\Delta V_{FG1}$, $\Delta V_{FG2}$, and tunnel coupling $\tau$ for a set of 500 defective electron (blue) and hole (green) Si MOS devices with trap density $n_i=5\times 10^{10}$\,cm$^{-2}$.}
    \label{fig:supp_scatter_V} 
\end{figure}

\begin{figure}
    \centering
    \includegraphics{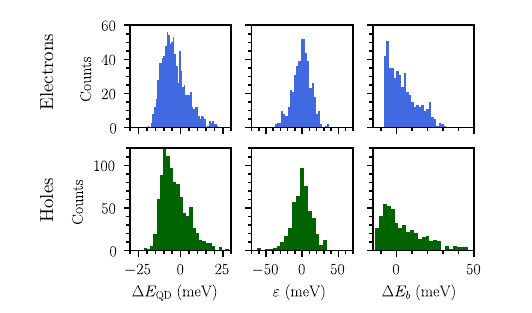}
    \caption{Distribution of the energy shifts $\Delta E_{\rm QD}$, $\varepsilon$ and $\Delta E_b$ for a set of 500 defective electron (blue) and hole (green) Si MOS devices with trap density $n_i=5\times 10^{10}$\,cm$^{-2}$.}
    \label{fig:supp_scatter_E} 
\end{figure}

The methodology we use for Si MOS devices is slightly different from that for Ge/GeSi heterostructures. The algorithm discussed in the main text indeed fails to bring many disordered devices back to the target tunnel coupling $\tau^0=10$\,$\mu$eV. The complex potential landscape resulting from the stronger disorder is much more difficult to explore (and would likely be so from an experimental point of view). Therefore, we simply tune the defective devices back to the reference chemical potential $\mu=\mu^0$ and to zero detuning energy $\varepsilon$ at constant $V_J$, and track $\Delta V_{FG1}$, $\Delta V_{FG2}$ as well as the gap $\Delta=2\tau$ at that anti-crossing. We then convert the latter into an effective barrier height variation $\Delta E_b$ using the $\tau(E_b)$ data calculated in the pristine device and displayed in Fig. \ref{fig:supp_device}b. No solution was found in $\approx 5$\% of the devices because there is actually no barrier (thus a single dot) at the corrected $V_{FG1}$, $V_{FG2}$ (but constant $V_J$).

The distributions of bias shifts in Si MOS devices are plotted in Fig.~\ref{fig:supp_scatter_V}. Contrarily to the Ge/SiGe heterostructures, the distributions are clearly non-normal, a fingerprint of the much larger impact of disorder. The standard deviation of the bias shifts is $\sigma(\Delta V_{FG1})=12.0$\,meV, $\sigma(\Delta V_{FG2})=10.6$\,meV for electrons, and $\sigma(\Delta V_{FG1})=12.4$\,meV, $\sigma(\Delta V_{FG2})=13.8$\,meV for holes. Nonetheless, the bias shifts in Si MOS and Ge/GeSi devices can hardly be compared, as the gate lever arms are totally different. Those of electrons and holes in FDSOI devices are given in Table \ref{tab:leverarms}. The electrostatic control by the front gates $P1$ and $P2$ is much tighter than in Ge/GeSi heterostructures owing to their close proximity to the QDs, yet the $J$ gates are far less efficient (because they are much farther and screened by the front gates). The comparison of the energy shifts $\Delta E_{\rm QD}$, $\varepsilon$ and $\Delta E_b$, which renormalize the lever arms, is more meaningful. The distributions of $\Delta E_{\rm QD}$, $\varepsilon$ and $\Delta E_b$ are plotted in Fig.~\ref{fig:supp_scatter_E}. Their standard deviations are $\sigma(\Delta E_{\rm QD})=7.4$\,meV, $\sigma(\varepsilon)=12.5$\,meV, and $\sigma(\Delta E_b)=9.3$\,meV for electrons, and $\sigma(\Delta E_{\rm QD})=8.8$\,meV, $\sigma(\varepsilon)=15.1$\,meV, and $\sigma(\Delta E_b)=13.4$\,meV for holes. These values are significantly larger than those reported in the main text for Ge/SiGe qubits. This highlights how harmful charge traps can be, and how one can effectively alleviate their impact by bringing them further away from the qubits. The variability is only slightly larger for holes than for electrons.

\bibliography{tips}

\end{document}